\documentclass[aip,amsmath,amssymb,reprint]{revtex4-1}

\usepackage{epsfig}
\usepackage{amsmath}

\def\bra #1 {\langle {#1} \vert}
\def\ket #1 {\vert {#1} \rangle}

\def\eb{\nonumber \\}

\usepackage{simplewick}
\usepackage{etoolbox}
\usepackage{color}

\makeatletter
\def\@email#1#2{%
 \endgroup
 \patchcmd{\titleblock@produce}
  {\frontmatter@RRAPformat}
  {\frontmatter@RRAPformat{\produce@RRAP{*#1\href{mailto:#2}{#2}}}\frontmatter@RRAPformat}
  {}{}
}%
\makeatother

\begin{document}

\title{Nonunitary projective transcorrelation theory inspired by the F12 ansatz}

\author{Seiichiro L. Ten-no}
\email{tenno@garnet.kobe-u.ac.jp}
\affiliation{Graduate School of System informatics, Kobe University, \\Nada-ku, Kobe 657-8501, Japan}

\date{\today}

\begin{abstract}
An alternative nonunitary transcorrelation, inspired by the F12 ansatz, is investigated.
In contrast to the Jastrow transcorrelation of Boys-Handy, the effective Hamiltonian of this projective transcorrelation features:\\
1. a series terminating formally at four-body interactions.\\
2. no spin-contamination within the non-relativistic framework.\\ 
3. simultaneous satisfaction of the singlet and triplet first-order cusp conditions.\\
4. arbitrary choices of pairs for correlation including frozen-core approximations.\\
We discuss the connection between the projective transcorrelation and F12 theory with applications to small molecules, to show that the cusp conditions play an important role to reduce the uncertainty arising from the nonunitary transformation.
\end{abstract}
\maketitle

\section{Introduction}\label{sec:intro}
Computing electronic structures in high accuracy is one of the most significant challenges in quantum chemistry and physics.
However, the exponential growth in the size of the Hilbert space, as the number of correlated electrons and orbitals increases, poses a hindrance to the application of wavefunction theory to large molecules.
A major obstacle lies in the cusp behavior that emerges at the coalescence of the exact wavefunction due to the Coulomb singularity, which requires the utilization of expensive orbital space.
Explicitly correlated electronic structure theory has alleviated this issue by incorporating ans\"atze that explicitly involve inter-electronic distances.
Particularly, F12 theory, building on the R12 technology initiated by the seminal work of Kuzelnigg,\cite{Kutzelnigg_1985} circumvents the explicit calculation of many-electron integrals.
Over the last two decades, it has rapidly evolved and grown.
Thanks to fundamental advancements, along with its integration with local correlation methods and high-performance computing, F12 theory has become a widely adopted standard tool, offering an accurate and robust treatment of complex systems.
Readers interested in the advancements of F12/R12 methods can refer to various review articles.\cite{Rev_KMTV2006,Rev_HKT2008,Rev_TN2012,Rev_T2012,Rev_HKKT2012,Rev_KBV2012,Rev_SW2013,Rev_GHOT2017,Rev_MW2018}

Another explicitly correlated approach is based on the similarity-transformed Hamiltonian
\begin{align}
\hat H_{\rm TC} &=e^{-J} \hat H e^J, \label{eq:htc}
\end{align}
with a Jastrow-type correlator $J=\sum^N_{i > j}j({\bf x}_i,{\bf x}_j)$ generally as a function of spatial-spin coordinates $\{ {\bf x}_i \}$.
As $j$ is commutable with $\hat H$ except for the kinetic energy operators $\hat t_i$, the expansion of $\hat H_{\rm TC}$ terminates at the double commutator involving the three-body operator $[[\hat t_i,j_{ij}],j_{ik}]$.
Hirshfelder advocated that the choice of $\exp(j_{ij})=1+\frac{1}{2}r_{ij}\exp(-\alpha r_{ij})$ removes the Coulomb pole $1/r_{ij}$ in the original Hamiltonian,\cite{Hirschfelder} in accord with Kato's cusp condition.\cite{Cusp_Kato}
The transcorrelated (TC) method of Boys and Handy follows this idea.\cite{Boys-Handy1,Boys-Handy2,Handy_TC}
One of the major difficulties arising from $\hat H_{\rm TC}$ is that the nonhermiticity sacrifices the variational upper-bound.
In other words, the obtained energy is not a measure to judge the accuracy of the solution.
Since the 2000s, there have been some developments in TC.
Practical perturbation and coupled-cluster theories with TC were proposed for polyatomic molecules, using frozen Gaussian geminals and the resolution of the identity (RI) approximation for three-electron integrals.\cite{ST_TC_2000a,ST_TC_2000b,Hino2001,Hino2002}
TC has been applied to homogeneous electron gas.\cite{Armour,Umezawa_2004,Luo}
Applications to solids have been intensively studied by Tsuneyuki and coworkers.\cite{Sakuma_2006,Ochi_2012,Ochi_2014a,Ochi_2014b}

While F12 theory remains the predominant method for accurate molecular electronic structures, there has been a recent resurgence in the interest of TC primarily driven by two objectives.
The first objective is to combine TC with approximate full configuration interaction (FCI) solvers like density matrix renormalization group, selected CI, and quantum Monte Carlo methods.\cite{cohen2019similarity,dobrautz2022performance,haupt2023optimizing,baiardi2020transcorrelated,baiardi2022explicitly,liao2023density,ammar2022optimization,ammar2022extension} 
The second is to reduce the number of qubits in quantum simulations by downfolding the high-energy contribution of the Hamiltonian.\cite{mcardle2020improving,sokolov2023orders}
For the latter purpose, the F12 canonical transcorrelation (F12-CT) approach of Yanai and Shiozaki\cite{yanai2012} is an alternative choice.\cite{motta2020quantum,schleich2022improving,kumar2022quantum}
F12-CT employs the unitary transformation based on the F12 ansatz inheriting all the advantages of F12 theory over methods employing Jastrow factors.
Especially, the flexibility of the formulation in the orbital space allows to use the singlet and triplet cusp conditions,\cite{Cusp_Kato,Cusp_Pack,Cusp_Kutzelnigg} which cannot be satisfied simultaneously by Jastrow factors.
However, on the negative side, the unitary transformation in F12-CT results in a nonterminating series, necessitating a drastic approximation to truncate the effective Hamiltonian expansion.

The objective of this paper is to investigate a nonunitary projective transcorrelation (pTC) in a terminating series.
The pTC formulation along with the connection to F12 theory will be presented in the following section.
A numerical analysis of pTC is performed perturbatively using small molecules in Sec. \ref{sec:results}, followed by conclusions in Sec. \ref{sec:conclusions}.

\section{Theory}\label{sec:theory}
\subsection{Formulation of pTC}
Throughout this paper, we use the conventions for orbital indices and projection operators in Table \ref{tab:index}, along with the Einstein summation convention unless indicated.
Let us consider the generator,
\begin{align}
\hat {\mathcal G} = \frac{1}{2} \langle \kappa \lambda |  \hat{\mathcal Q}_{12} \hat {\mathcal R}_{12} | rs\rangle \hat E_{rs}^{\kappa\lambda}, \label{genetator}
\end{align}
where $\hat{\mathcal Q}_{12}$ is a suitable orthogonalization operator to project out the geminal contributions representable by the given basis set (GBS) from those in the complete basis set (CBS),
\begin{align}
\hat {\mathcal Q}_{12} &=  (1-\hat P_1\hat P_2) \hat {\mathcal S}_{12}, \label{eq:projector}
\end{align}
$\hat {\mathcal S}_{12}$ is a supplementary operator, $\hat {\mathcal R}_{12}$ is the rational generator,\cite{ST_JCP_2004}
\begin{align}
\hat {\mathcal R}_{12} =  \{g^{(d)}({\bf r}_{1},{\bf r}_{2}) + g^{(x)}({\bf r}_{1},{\bf r}_{2}) \hat {\mathcal P}_{12}\},
\end{align}
with a permutation operator $\hat {\mathcal P}_{12}$ to interchange the spatial functions, e.g. $\hat {\mathcal P}_{12} | pq \rangle=| qp \rangle$, and $\hat E^{\kappa\lambda\dots}_{\mu\nu\dots}=\hat a^{\dagger}_{\kappa\sigma_1}\hat a^{\dagger}_{\lambda\sigma_2}\dots \hat a_{\nu\sigma_2}\hat a_{\mu\sigma_1}$ are spin-free orbital replacement operators.
Unlike the Jastrow factor, the rational generator with the asymptotic behaviors,
\begin{align}
g^{(d)}({\bf r}_{1},{\bf r}_{2}) &= \frac{3}{8}r_{12}+O(r^2_{12}) \\
g^{(x)}({\bf r}_{1},{\bf r}_{2}) &= \frac{1}{8}r_{12}+O(r^2_{12}),
\end{align}
satisfies the s- and p-wave cusp conditions simultaneously.\cite{ST_JCP_2004}
The simplest variant, the so-called SP or fixed amplitudes (FIX) ansatz,
\begin{align}
\hat {\mathcal R}_{12} = f(r_{12})(\frac{3}{8}+\frac{1}{8}\hat {\mathcal P}_{12}), \label{sp_ansatz}
\end{align}
in combination with the Slater-type correlation factor,\cite{ST_CPL_2004}
\begin{align}
f(r_{12}) = -\frac{1}{\gamma} \exp(-\gamma r_{12}),
\end{align}
is diagonal orbital invariant (DOI), without linear dependencies, free from geminal superposition errors,\cite{GBSSE} and has facilitated the popularity of F12 theory.

\begin{table}[t]
\caption
{\label{tab:index}
Convention of orbital indices and projectors.}
\begin{tabular}{lcc}
\hline
Indices & Type of Orbitals & Projector \\
\hline
$p,q, ...$ & Orbitals in the GBS & $\hat P_{n}$ \\
$\alpha, \beta, ...$ & Orthogonal complement to GBS & $\hat Q_{n}$ \\
$\kappa, \lambda, ...$ & Orbitals in the CBS & $1=\hat P_{n}+\hat Q_{n}$ \\
\hline
$i,j, ...$ & Occupied orbitals & $\hat O_{n}$ \\
$a,b, ...$ & Virtual orbitals & $\hat V_{n}$ \\
$i', j', ...$ & Core orbitals in the occupied space & $\hat O'_{n}$ \\
$v', w', ...$ & Valence orbitals in the occupied space & $\hat O_{n}-\hat O'_{n}$ \\
\hline
\end{tabular}
\end{table}

The rational generator $\hat {\mathcal R}_{12}$ commutes with any potentials in the Hamiltonian as Jastrow factors do; however, the similarity transformation $e^{-\hat {\mathcal R}}\hat H e^{\hat {\mathcal R}}$ becomes nonterminating due to $\hat {\mathcal P}_{12}$.\cite{ST_JCP_2004}
Thus, the projector form in Eq. (\ref{eq:projector}) is essential to realize a terminating expression of pTC. 
If a partitioning between occupied and virtual orbitals is clear in GBS as in the single reference (SR) case, the natural choice of $\hat {\mathcal S}_{12}$ is the strong-orthogonality (SO) projector,\cite{Klopper_JPC_1990}
\begin{align}
\hat {\mathcal S}^{\rm (SO)}_{12} &= (1-\hat O_1)(1-\hat O_2) \\
\hat {\mathcal Q}^{\rm (SO)}_{12} &= 1-\hat O_1\hat Q_2 - \hat Q_1\hat O_2 -\hat P_1\hat P_2, \label{SOP}
\end{align}
sometimes called modified ansatz 2 (or ansatz 3)\cite{Werner2007} ensuring the orthogonality within the Hilbert space spanned by GBS, $\langle \Phi_I | \hat {\mathcal G} | \Phi_J \rangle = 0$, and those between the reference determinant $\Phi$ and all singles in the complementary space, $\langle \hat E^{\alpha}_i \Phi | \hat {\mathcal G} | \Phi \rangle = 0$.
The resulting ${\mathcal G}$ resembles the extended SP ansatz of K\"ohn for response properties.\cite{kohn2009modified} 
The same projector was also employed in the multireference (MR) framework for the F12 treatment of external excitations.\cite{ST_CPL_2007,Shio2010}
However, SO is not suitable when the occupied space cannot be defined uniquely.
Another choice is the universal (U) projector that just excludes the contribution of core orbitals,
\begin{align}
\hat {\mathcal S}^{(\rm U)}_{12}&=(1-\hat O'_1)(1-\hat O'_2) \\
\hat {\mathcal Q}^{(\rm U)}_{12} &=  1-\hat O'_1\hat Q_2 - \hat Q_1\hat O'_2 -\hat P_1\hat P_2.\label{eq:UP}
\end{align}
For all-electron correlation, this reduces to $\hat {\mathcal Q}^{(\rm U)}_{12} =  1-\hat P_1\hat P_2$, initially introduced to approximate SO within the standard approximation (SA),\cite{KK1} and later employed by Torheyden and Valeev for a universal F12 perturbative correction.\cite{torheyden2009universal}
We do not introduce any normal ordering to $\hat {\mathcal G}$ for the sake of universality of pTC.
In this case, $\hat {\mathcal Q}^{(\rm U)}_{12}$ also ensures the orthogonality within GBS, $\langle \Phi_I | \hat {\mathcal G} | \Phi_J \rangle = 0$, but not for complementary singles, $\langle \hat E^{\alpha}_{v' }\Phi | \hat {\mathcal G} | \Phi \rangle \ne 0$.
However, this contribution is considered to vanish quickly with the size of GBS, e.g. the angular momentum of $\alpha$ for nonzero $\langle \hat E^{\alpha}_{v'} \Phi | \hat {\mathcal G} | \Phi \rangle \ne 0$ is bound at most at $3L_{\rm occ.}$ for the maximum angular momentum of the occupied orbitals $L_{\rm occ.}$ in $\Phi$.

We now define the pTC Hamiltonian as the similarity transformation with $\hat {\mathcal G}$,
\begin{align}
\hat {\mathcal H}_{\rm pTC} &= {\mathcal B}\{ e^{-\hat {\mathcal G}} \hat H e^{\hat {\mathcal G}} \},
\end{align}
where ${\mathcal B}\{\dots\}$ denotes the second quantized representation within GBS.
$\hat {\mathcal G}$ contains at least one creation operator in the complementary space, which should be contracted with $\hat H$ via $\hat Q_{n}$ for nonzero contributions to $\hat {\mathcal H}_{\rm pTC}$.
The $\hat {\mathcal H}_{\rm pTC}$ then terminates at the double commutator,
\begin{align}
\hat {\mathcal H}_{\rm pTC} &={\mathcal B}\{ \hat H +[\hat H, \hat {\mathcal G}]+\frac{1}{2}[[\hat H, \hat {\mathcal G}],\hat {\mathcal G}]\} \eb
&= \hat {\mathcal H} + \hat {\mathcal H}_{\rm 2h} + \hat {\mathcal H}_{\rm 2w} + \hat {\mathcal H}_{\rm 3l} + \hat {\mathcal H}_{\rm 3q} + \hat {\mathcal H}_{4}, \label{eq:hpTC}
\end{align}
where $\hat {\mathcal H}= {\mathcal B}\{\hat H\}$, and the contractions through four-body operators in skeleton forms are
\begin{align}
&\hat {\mathcal H}_{\rm 2h} = {\mathcal B}\{\contraction{}{H}{_1}{\hat {\mathcal G}} \hat H_1 \hat {\mathcal G} \} = \langle \hat h_{1} \hat Q_1 \hat {\mathcal S}_{12} \hat {\mathcal R}_{12} \rangle^{rs}_{pq} \hat E^{pq}_{rs} \\
&\hat {\mathcal H}_{\rm 2w} = {\mathcal B}\{\contraction{}{H}{_2}{\hat {\mathcal G}} \contraction[2ex]{}{H}{_2}{\hat {\mathcal G}} \hat H_2 \hat {\mathcal G}\} = \frac{1}{2} \langle r^{-1}_{12} \hat {\mathcal Q}_{12} \hat {\mathcal R}_{12} \rangle^{rs}_{pq} \hat E^{pq}_{rs} \\
&\hat {\mathcal H}_{\rm 3l} = {\mathcal B}\{\contraction{}{H}{_2}{\hat {\mathcal G}} \hat H_2 \hat {\mathcal G} \} = \langle r^{-1}_{12} \hat Q_1 \hat {\mathcal S}_{13}  \hat {\mathcal R}_{13} \rangle^{stu}_{pqr} \hat E^{pqr}_{stu} \\
&\hat {\mathcal H}_{\rm 3q} = {\mathcal B}\{\frac{1}{2}\contraction{}{H}{_2}{\hat {\mathcal G}} \contraction[2ex]{}{H}{_2\hat {\mathcal G}}{\hat {\mathcal G}} \bcontraction{H_2}{{\mathcal G}}{}{\hat {\mathcal G}} \hat H_2 \hat {\mathcal G} \hat {\mathcal G} \}= \langle r^{-1}_{12} \hat Q_1\hat Q_2 \hat {\mathcal U}_{123} \rangle^{stu}_{pqr}
\hat E^{pqr}_{stu} \\
&\hat {\mathcal H}_{4} = {\mathcal B}\{\frac{1}{2}\contraction[2ex]{}{H}{_2\hat {\mathcal G}}{\hat {\mathcal G}} \contraction{}{H}{_2}{\hat {\mathcal G}} \hat H_2 \hat {\mathcal G} \hat {\mathcal G} \} = \langle r^{-1}_{12} \hat Q_1\hat Q_2 \hat {\mathcal W}_{1234} \rangle^{tuvw}_{pqrs} \hat E^{pqrs}_{tuvw},
\end{align}
the lines connecting operators denote Wick contractions of ordinary products, the short-hand notation like $\langle \dots \rangle ^{rs}_{pq}=\langle pq |\dots |rs \rangle$ is used, and $\hat {\mathcal U}_{123}$ and $\hat {\mathcal W}_{1234}$ are defined as
\begin{align}
&\hat {\mathcal U}_{123} = \frac{1}{2}\hat {\mathcal S}_{23} \hat {\mathcal R}_{23} (\hat P_2 \hat {\mathcal S}_{12} \hat {\mathcal R}_{12}+ \hat P_3 \hat {\mathcal S}_{13} \hat {\mathcal R}_{13}) \\
&\hat {\mathcal W}_{1234} = \frac{1}{2} \hat {\mathcal S}_{13} \hat {\mathcal R}_{13} \hat {\mathcal S}_{24} \hat {\mathcal R}_{24}.
\end{align}
It is straightforward to obtain the matrix elements using the generalized Wick theorem.\cite{mukherjee2012aspects}
$\hat h_1$ contains the frozen-core contribution and the core indices are excluded from $\hat {\mathcal H}_{\rm pTC}$ for the frozen-core approximation.
Among the operators, $\hat {\mathcal H}_{\rm 2h}$ and $\hat {\mathcal H}_{\rm 2w}$, play an essential role for the fast convergence.
With respect to the leading terms of singlet and triplet pairs, $r^0_{12}$ and $r^1_{12}$, respectively, the Coulomb singularity $r^{-1}_{12}|rs \rangle$ of $\hat {\mathcal H}$ is regularized by the kinetic energy operator in $\hat {\mathcal H}_{\rm 2h}$ as
\begin{align}
(\hat t_{1}+\hat t_{2}) \hat {\mathcal R}_{12}|rs \rangle &= \{-r^{-1}_{12}+O(r^0_{12})\} |rs \rangle,
\end{align}
both for $r+s$ (singlet) and $r-s$ (triplet).\cite{Rev_T2012}
Instead, $\hat {\mathcal H}_{\rm 2w}$ represents the residual correlation in the complementary space.
The integrals in the above expressions can be computed either by using (complementary) auxiliary basis set ((C)ABS),\cite{Klopper2002,Valeev2004a} or numerical integrations\cite{ST_JCP_2004,ST_JCP_2007} after the decompositions of $\hat Q_n=1-\hat P_n$, $\hat {\mathcal Q}_{12}$, and $\hat {\mathcal S}_{12}$, with two-electron objects for $1/r_{12}$, $f(r_{12})$, and $f(r_{12})/r_{12}$.
The two-body operators quadratic to the correlation factor, $f^2(r_{12})$ and $\nabla f(r_{12}) \cdot \nabla f(r_{12})$, in the usual F12 methods do not appear in pTC.

\subsection{Connection to the F12 wavefunction}
$\hat {\mathcal H}_{\rm pTC}$ gives the same energy spectrum of the original Hamiltonian $\hat H$ in CBS.
We look into $\hat {\mathcal H}_{\rm pTC}$ pertubatively using the Hartree-Fock (HF) reference with the modified M{\o}ller-Plesset (MP) partitioning,
\begin{align}
\hat {\mathcal H}_{\rm pTC} &= \hat {\mathcal H}_{0} + \hat {\mathcal V}_{\rm pTC} \\
\hat {\mathcal V}_{\rm pTC} &= \hat {\mathcal V} + \hat {\mathcal H}_{\rm 2h} + \hat {\mathcal H}_{\rm 2w} + \hat {\mathcal H}_{3} + \hat {\mathcal H}_{4},
\end{align}
with $\hat {\mathcal H}_{0}={\mathcal B}\{\hat H_{0}\}$ and $\hat {\mathcal V}={\mathcal B}\{\hat V\}$.
The second order energy of MP2-F12 with the same generator $\hat {\mathcal G}$ is derived from the Hylleraas energy functional,
\begin{align}
E^{(2)}=\langle \Psi^{(1)}|\hat H_{\rm 0N}| \Psi^{(1)} \rangle +2 \langle \Psi^{(1)}|\hat V| \Phi \rangle \ge E^{(2)}_{\infty},
\end{align}
where $\hat H_{\rm 0N}=\hat H_0-E^{(0)}$, $E^{(2)}_{\infty}$ is the exact second order energy, and $\Psi^{(1)}$ is a sum of the contribution from GBS and explicitly correlated part,
\begin{align}
|\Psi^{(1)} \rangle = |\Psi_{\rm O}^{(1)} \rangle + \hat {\mathcal G} | \Phi \rangle.
\end{align}
Minimizing $E^{(2)}$ with respect to $\Psi_{\rm O}^{(1)}$ yields
\begin{align}
E^{(2)}=E_{\rm MP2}+B+2V,
\end{align}
with the conventional correlation energy, $E_{\rm MP2}$, and F12 terms,
\begin{align}
B&= \langle \hat {\mathcal G}^{\dagger} \hat H_{\rm 0N} \hat {\mathcal G} - \hat {\mathcal G}^{\dagger} \hat H_{\rm 0N} \hat {\mathcal H}_{\rm 0N}^{-1}\hat H_{\rm 0N} \hat {\mathcal G} \rangle  \\
V&= \langle \hat V \hat {\mathcal G} - \hat V \hat {\mathcal H}_{\rm 0N}^{-1}\hat H_{\rm 0N} \hat {\mathcal G} \rangle,
\end{align}
where $\langle \hat A \rangle=\langle\Phi| \hat A |\Phi\rangle$, and $\hat {\mathcal H}_{\rm 0N}^{-1}$ means the MP preconditioner within the excitation manifold of GBS.
We do not repeat the derivation of detailed expressions for MP2-F12, which are available in the literature.\cite{Klopper2002,ST_JCP_2007,Werner2007}
\begin{table}[t]
\caption
{\label{tab:order}
Operators in each order of perturbation.}
\begin{tabular}{c|lcc}
\hline
Order & \multicolumn{3}{c}{Operators} \\
\hline
1 & $\hat {\mathcal V}$ & $\hat {\mathcal H}_{\rm 2h}$ & $\hat {\mathcal H}_{\rm 2v}$ \\
2 & $\hat {\mathcal H}_{\rm 2w}$ & \multicolumn{1}{c}{$[\hat {\mathcal H}_{\rm 3l}]_{\rm r}$} &\\
3 & $\hat {\mathcal H}_{\rm 3q}$ & \multicolumn{1}{c}{$\hat {\mathcal H}_{4}$} &\\ 
\hline
\end{tabular}
\end{table}
For later discussions, $\hat {\mathcal H}_{\rm 3l}$ is divided into the contribution of the HF mean field, and its residual,
\begin{align}
\hat {\mathcal H}_{\rm 3l}&=\hat {\mathcal H}_{\rm 2v} + [\hat {\mathcal H}_{\rm 3l}]_{\rm r} \\
\hat {\mathcal H}_{\rm 2v} &= \langle pq | \hat v^{\rm (HF)}_{1} \hat Q_1 \hat {\mathcal S}_{12} \hat {\mathcal R}_{12}|rs\rangle \hat E^{pq}_{rs},
\end{align}
and the contraction with the Fock operator, $\hat f_{1}=\hat h_{1}+\hat v^{\rm (HF)}_{1}$, is defined as
\begin{align}
\hat {\mathcal H}_{\rm 2f} = \hat {\mathcal H}_{\rm 2h} + \hat {\mathcal H}_{\rm 2v}.
\end{align}
This division is unique in the SR case, but not in general.
Table \ref{tab:order} summarizes the orders of the operators.
At first glance of the expectation value $\langle \hat {\mathcal V}_{\rm pTC} \rangle$, any components in $B$, which are quadratic to $\hat {\mathcal G}$, are absent.
$\langle \hat {\mathcal H}_{\rm 3q} \rangle$ and $\langle \hat {\mathcal H}_{4} \rangle$ are third-order contributions from $\frac{1}{2}\hat {\mathcal G}^2$ with different operator alignments.
Actually, besides the higher-order corrections from $\hat {\mathcal H}_{\rm 3q}$ and $\hat {\mathcal H}_{4}$, $\langle \hat {\mathcal V}_{\rm pTC} \rangle$ is identified with $V$ under the assumption of the extended Brillouin condition (EBC),\cite{Klopper2002} $[\hat P_{1},\hat f_{1}]= 0$, plus the first-order contribution of $\langle \hat {\mathcal H}_{\rm 2f} \rangle$ not present in the standard MP2-F12 due to $\hat {\mathcal G}^{\rm (SO)}$ and the assumption of the generalized Brillouin condition (GBC),\cite{KK1} $[\hat O_{1},\hat f_{1}]= 0$.
The correction term to EBC arises from the second order energy, $\langle -\hat {\mathcal V} \hat {\mathcal H}_{\rm 0N}^{-1} \hat {\mathcal H}_{\rm 2f}\rangle$ (cf. Eq. (28) in Ref. \onlinecite{ST_JCP_2007}) by the relation,
\begin{align}
\langle ab|\hat F_{12} \hat {\mathcal Q}_{12} \hat {\mathcal R}_{12}|ij\rangle &=\langle ab|(\hat F_{12}-\epsilon_i-\epsilon_j)\hat {\mathcal R}_{12}|ij\rangle \eb
&- \langle ab|\hat {\mathcal R}_{12} |ij\rangle \Delta_{ab,ij}, \label{correctEBC}
\end{align}
where we use the notations,
\begin{align}
\hat F_{12}&=\hat f_{1}+\hat f_{2} \\
\Delta_{ab,ij}&=\epsilon_a+\epsilon_b-\epsilon_i-\epsilon_j.
\end{align}
$\langle -\hat {\mathcal H}_{\rm 2f} \hat {\mathcal H}_{\rm 0N}^{-1} \hat {\mathcal V}\rangle$ is also absent in the standard MP2-F12 due to the GBC.
In this way, $\langle \hat {\mathcal H}_{\rm pTC} \rangle$ contains no contributions in the B-term of MP2-F12.
However, {provided a near-optimum generator $\hat {\mathcal G}$ minimizing $E^{(2)}$ is available, the approximation, $V\approx -B$, holds and thus
\begin{align}
E^{(2)} \overset{\text{opt}}{\approx} E_{\rm MP2}+V.
\end{align}
This approximation was employed in solid state calculations by Gr\"uneis.\cite{Gruneis2015}
Nevertheless, a variational optimization of $\hat {\mathcal G}$ again requires the integrals over $f^2(r_{12})$ and $\nabla f(r_{12}) \cdot \nabla f(r_{12})$ within the F12 framework, which has limited the flexibility in the choice of $f(r_{12})$ or $f({\bf r}_{1},{\bf r}_{2})$.

Higher-order F12 contributions can be automatically included by plugging $\hat {\mathcal H}_{\rm pTC}$ into correlated calculations.
$\langle -\hat {\mathcal V} \hat {\mathcal H}_{\rm 0N}^{-1} (\hat {\mathcal H}_{\rm 2w}+\hat {\mathcal H}_{\rm 3l}) \rangle$ corresponds to the MP3(F12) correction\cite{Haettig2012,ohnishi2013alternative} as the leading term of approximate CCSD-F12 models.\cite{fliegl2005coupled,tew2007quintuple,Adler2007,torheyden2008variational,Knizia2009,bokhan2009implementation}
The extended SP ansatz of K\"ohn\cite{kohn2009explicitly} for explicitly correlated connected triples is included as the contraction between $\hat {\mathcal H}_{\rm pTC}$ and $\hat T_{2}$ in CCSD with $\hat {\mathcal H}_{\rm pTC}$.
Other contributions from the three- and four-electron operators, $\hat {\mathcal H}_{\rm 3q}$ and $\hat {\mathcal H}_{4}$, containing two $\hat {\mathcal R}_{12}$ connected with $1/r_{12}$ through $\hat Q_n$ is considered to be very small and usually neglected in F12 theory.

\subsection{Hybrid ansatz for core-valence correlation}
The SP ansatz is the generalization of the linear $r_{12}$ treatment of He-like ions by Kutzelnigg with fixed geminal coefficients,\cite{Kutzelnigg_1985}
and the association with the Slater-type factor usually gives very accurate results.
It is intended to employ core functions for core-valence correlation effects, since basis functions in lower angular momentum are presumed to be saturated for fast CI convergence of partial wave expansions with the r12 behavior.\cite{Kutzelnigg_1985}
cc-pCV{\it n}Z-F12 basis sets\cite{hill2010correlation,hill2010correlation2} have been developed with this objective, with fewer augmented functions compared to the standard cc-pCV{\it X}Z basis sets.\cite{Dunning1989,kendall1992electron,Woon1995}
However, the SP ansatz does not show a sufficient ability for core-valence energies with an ill-matched basis set.
Optimizing geminal amplitudes can improve this problem, but introduces other difficulties.
The orbital-invariant (OI) ansatz\cite{Klopper1991} for four-index amplitudes is known to suffer from linear-dependency problem in the B-intermediate.
The diagonal (D) ansatz is free from linear-dependency, violating the orbital invariance.
The pair-specific geminal approach\cite{werner2011explicitly} and relaxing constrained amplitudes method\cite{tew2018relaxing} have been proposed to improve upon this situation.

In this work, we introduce the hybrid (HYB) ansatz, which adopts different ans\"atze for different orbital pairs, i.e. SP, D, and OI for valence-valence (VV), core-valence (CV), and core-core (CC) pairs, respectively,
\begin{align}
&\hat {\mathcal G}^{\rm (HYB)} = \hat {\mathcal G}_{\rm VV}+\hat {\mathcal G}_{\rm CV}+\hat {\mathcal G}_{\rm CC}\\
&\hat {\mathcal G}_{\rm VV}= \frac{1}{2} \langle \kappa \lambda |  \hat{\mathcal Q}_{12} \hat {\mathcal R}_{12} | v'w'\rangle \hat E_{v'w'}^{\kappa\lambda} \\
&\hat {\mathcal G}_{\rm CV}= \langle \kappa \lambda |  \hat{\mathcal Q}_{12} f(r_{12}) (| i'v'\rangle t'^{i'v'}_{i'v}+| v'i'\rangle t'^{i'v'}_{v'i'})\hat E_{i'v'}^{\kappa\lambda} \\
&\hat {\mathcal G}_{\rm CC}= \frac{1}{2} \langle \kappa \lambda |  \hat{\mathcal Q}_{12} f(r_{12}) | k'l'\rangle t'^{k'l'}_{i'j'} \hat E_{i'j'}^{\kappa\lambda},
\end{align}
where $t'^{i'v'}_{i'v'}$ and $t'^{i'v'}_{v'i'}$ are geminal amplitudes for the CV block, and $t'^{k'l'}_{i'j'}$ are those for CC.
The HYB ansatz is orbital invariant with respect to the transformations within the core and valence spaces separately, and the drawbacks of OI in the scaling and linear dependency are substantially reduced owing to the application within CC.
OI requires cumbersome B-term intermediates with different orbital energies in numerators and denominators.
We avoid this problem using the geminal amplitudes of MP2-F12 under the EBC, and the correction to the EBC approximation is added in the V-term similarly to Eq. (\ref{correctEBC}).
We examine the performance of the HYB adaptation of pTC in the following section.

\section{RESUlTS AND DISCUSSION}\label{sec:results}
We investigate pTC numerically by comparing the second order contribution of $\hat {\mathcal H}_{\rm pTC}$ (pTC(2)) with variational MP2-F12 for the selection of small closed-shell molecules, CH$_2$($^1$A$_1$), H$_2$O, NH$_2$, HF, N$_2$, CO, Ne, and F$_2$, employed in the previous assessments,\cite{Klopper2002,ST_JCP_2004,ST_JCP_2007} with aug-cc-p(C)V{\it X}Z basis sets.\cite{Dunning1989,kendall1992electron,Woon1995}
The augmentation with diffuse functions is not essential for discussions but improves the long-range correlation that is not universal.\cite{ST_JCP_2004}
The present pTC(2) implementation resembles that for MP2-F12.\cite{ST_JCP_2004,ST_JCP_2007}
All necessary integrals are calculated using numerical integrations along with extra operations} over two- and three-index objects (QD2 scheme in MP2-F12)\cite{ST_JCP_2004} containing extra operations for $\hat t _1 \phi_p({\bf r}_1)$ and $\hat v_1 \phi_p({\bf r}_1)$ with the ultra-fine grid of our new Lebedev implementation in the \texttt{GELLAN} program package\cite{gellan} resulting in 9,340 integration points for hydrogen and 23,842 points for the other atoms before imposing the point group symmetry.
The three-index integrals for $f(r_{12})$, and $f(r_{12})/r_{12}$ are computed from the generalized Boys function, $G_m(T,U)$.\cite{ST_CPL_2004,ST_JCP_2007}
The latest version of subroutines for $G_m(T,U)$ contain a minor revision for $T>20$ and $U>200$ with the downward recurrence relation starting with marginal $m$ to improve the accuracy.
Other implementations for Slater integrals with Chebyshev interpolation can be found elsewhere.\cite{Shiozaki_Int_2009,valeevlibint}
The Slater exponent $\gamma=1.5$ is used throughout this work.
Since pTC(2) does not require integrals for the B-term intermediates, the computational cost is almost half that of MP2-F12.

\begin{table}[t]
\begin{center}
\caption
{\label{tab:val_water}
Energy components of pTC(2) in mE$_{\rm h}$ for the valence-correlation energy of H$_2$O in the aug-cc-pV{\it X}Z basis sets. The MP2 limit is ca. -300.5 mE$_{\rm h}$.}
\scalebox{1.00}{
\begin{tabular}{lrrrrr}
\hline\hline
& \multicolumn{5}{c}{{\it X}} \\
\cline{2-6}
& \multicolumn{1}{c}{D} & \multicolumn{1}{c}{T} & \multicolumn{1}{c}{Q} & \multicolumn{1}{c}{5} & \multicolumn{1}{c}{6} \\ 
\hline
1 $\langle \hat {\mathcal H}_{\rm 2h} \rangle$ & 19.13 & 1.00 &  0.21 & 0.01 & 0.01\\
2 $\langle \hat {\mathcal H}_{\rm 2v} \rangle$ & -16.97 & -0.39 & -0.11 & 0.00 & 0.00 \\
3 $\langle \hat {\mathcal H}_{\rm 2w} \rangle$ & -79.11 & -31.88 & -14.27 & -7.46 & -4.46 \\
4 $\langle [\hat {\mathcal H}_{\rm 3l}]_{\rm r} \rangle$ & 12.98 & 2.13 & 0.22 & 0.04 & 0.01 \\
5 $\langle -\hat {\mathcal V} \hat {\mathcal H}^{-1}_{\rm 0N} \hat {\mathcal H}_{\rm 2h}  \rangle$ & -3.27 & -0.56 & -0.20 & -0.02 & -0.02 \\
6 $\langle -\hat {\mathcal V} \hat {\mathcal H}^{-1}_{\rm 0N} \hat {\mathcal H}_{\rm 2v} \rangle$ &  4.60 &  0.86 & 0.28 & 0.07 & 0.02 \\
7 $\langle -\hat {\mathcal H}_{\rm 2h}  \hat {\mathcal H}^{-1}_{\rm 0N} \hat {\mathcal V} \rangle$ & -1.66 &  -0.02 & -0.01 & 0.00 & 0.00 \\
8 $\langle -\hat {\mathcal H}_{\rm 2v} \hat {\mathcal H}^{-1}_{\rm 0N} \hat {\mathcal V}  \rangle$ &  1.61 &  0.02 & 0.01 & 0.00 & 0.00 \\
9 $\langle -\hat {\mathcal H}_{\rm 2f} \hat {\mathcal H}^{-1}_{\rm 0N} \hat {\mathcal H}_{\rm 2f}  \rangle$ & 0.00 & 0.00 & 0.00 & 0.00 & 0.00\\
SUM & -62.67 & -28.83 & -13.88  & -7.37 & -4.45 \\
MP2 & -219.35 & -268.36 & -285.93 & -292.92 & -295.97 \\
\hline
pTC(2) & -282.02 & -297.19 & -299.80 & -300.29 & -300.42 \\
$\Delta$ & 18.48 & 3.31 & 0.70 & 0.21 & 0.08 \\
\hline\hline
\end{tabular}}
\end{center}
\end{table}
We first investigate the energy components of pTC(2).
Tables \ref{tab:val_water} and \ref{tab:all_water} list the basis set dependency of these components along with the deviation from the exact second order energy,
\begin{align}
\Delta=E^{(2)}-E^{(2)}_{\infty},
\end{align}
for valence- and all-electron correlation energies of H$_2$O in aug-cc-pV{\it X}Z and aug-cc-pCV{\it X}Z basis set, respectively.
The sum of the terms 3 and 4 corresponds to the V-term of MP2-F12 under EBC, and that of 5 and 6 is equivalent to the correction to EBC.
The other terms absent in the standard MP2-F12 arise from the use of $\hat {\mathcal Q}^{(\rm U)}_{12}$,
$\langle -\hat {\mathcal H}_{\rm 2f}  \hat {\mathcal H}^{-1}_{\rm 0N} \hat {\mathcal H}_{\rm 2f} \rangle$ is negligibly small overall.
The others possess large contributions in aug-cc-p(C)VDZ and decay very quickly as the cardinal number {\it X} increases except for $\langle \hat {\mathcal H}_{\rm 2w} \rangle$ complementing the slow convergence of MP2.
$\langle \hat {\mathcal H}_{\rm 2h} \rangle$ and $\langle \hat {\mathcal H}_{\rm 2v} \rangle$ include the contractions containing $\hat h_1$ and $\hat v^{\rm (HF)}_1$ that cancel each other under the GBC, $\hat Q_1 (\hat h_1+ \hat v^{\rm (HF)}_1) | i \rangle =0$, usually employed in F12 methods.
Similar cancelation from GBC takes place between $\langle -\hat {\mathcal H}_{\rm 2h}  \hat {\mathcal H}^{-1}_{\rm 0N} \hat {\mathcal V} \rangle$ and $\langle -\hat {\mathcal H}_{\rm 2v}  \hat {\mathcal H}^{-1}_{\rm 0N} \hat {\mathcal V} \rangle$, albeit each contribution is much smaller.
For the general use of pTC, it is essential to argue each convergence of the terms since such cancellation is unavailable.
The rapid convergence of $\langle \hat {\mathcal H}_{\rm 2v} \rangle$ indicates the contributions of the third order operators, $\hat {\mathcal H}_{\rm 3q}$ and $\hat {\mathcal H}_{\rm 4}$, quadratic to $\hat Q_n$ attenuate even faster in correlation energies from the third order, and can be omitted especially for $X \ge {\rm T}$.

\begin{table}[t]
\begin{center}
\caption
{\label{tab:all_water}
Energy components of pTC(2) in mE$_{\rm h}$ for all-electron calculations of H$_2$O in the aug-cc-pCV{\it X}Z basis sets. The MP2 limit is ca. -362.1 mE$_{\rm h}$.}
\scalebox{1.00}{
\begin{tabular}{lrrrrr}
\hline\hline
& \multicolumn{5}{c}{{\it X}} \\
\cline{2-6}
& \multicolumn{1}{c}{D} & \multicolumn{1}{c}{T} & \multicolumn{1}{c}{Q} & \multicolumn{1}{c}{5} & \multicolumn{1}{c}{6} \\ 
\hline
1 $\langle \hat {\mathcal H}_{\rm 2h} \rangle$ & 110.47 & 3.06 & 0.38 & 0.04 & 0.01\\
2 $\langle \hat {\mathcal H}_{\rm 2v} \rangle$ & -106.74 & -2.20 & -0.16 & -0.01 & -0.01 \\
3 $\langle \hat {\mathcal H}_{\rm 2w} \rangle$ & -129.38 & -36.77 & -15.61 & -8.07 & -4.79 \\
4 $\langle [\hat {\mathcal H}_{\rm 3l}]_{\rm r} \rangle$ & 39.02 & 2.08 & 0.10 & 0.01 & 0.00 \\
5 $\langle -\hat {\mathcal V} \hat {\mathcal H}^{-1}_{\rm 0N} \hat {\mathcal H}_{\rm 2h}  \rangle$ & -10.94 & -2.43 & -0.72 & -0.28 & -0.14 \\
6 $\langle -\hat {\mathcal V} \hat {\mathcal H}^{-1}_{\rm 0N} \hat {\mathcal H}_{\rm 2v} \rangle$ & 7.59 & 0.97 & 0.26 & 0.07 & 0.02 \\
7 $\langle -\hat {\mathcal H}_{\rm 2h}  \hat {\mathcal H}^{-1}_{\rm 0N} \hat {\mathcal V} \rangle$ & -3.92 & -0.09 & -0.01 & 0.00 & 0.00 \\
8 $\langle -\hat {\mathcal H}_{\rm 2v} \hat {\mathcal H}^{-1}_{\rm 0N} \hat {\mathcal V}  \rangle$ & 3.89 & 0.10 & 0.00& 0.00 & 0.00 \\
9 $\langle -\hat {\mathcal H}_{\rm 2f} \hat {\mathcal H}^{-1}_{\rm 0N} \hat {\mathcal H}_{\rm 2f}  \rangle$ & -0.02 & 0.00 & 0.00& 0.00 & 0.00 \\
SUM & -90.02 & -35.29 & -15.76 & -8.26 & -4.91 \\
MP2 & -259.25 & -324.15 & -345.65 & -353.65 & -357.12 \\
\hline
pTC(2) & -349.27 & -359.44 & -361.41 & -361.91 & -362.03 \\
$\Delta$ & 12.83 & 2.66 & 0.69 & 0.19 & 0.07 \\
\hline\hline
\end{tabular}}
\end{center}
\end{table}
\begin{figure}[b]
	\includegraphics[width = 26em]{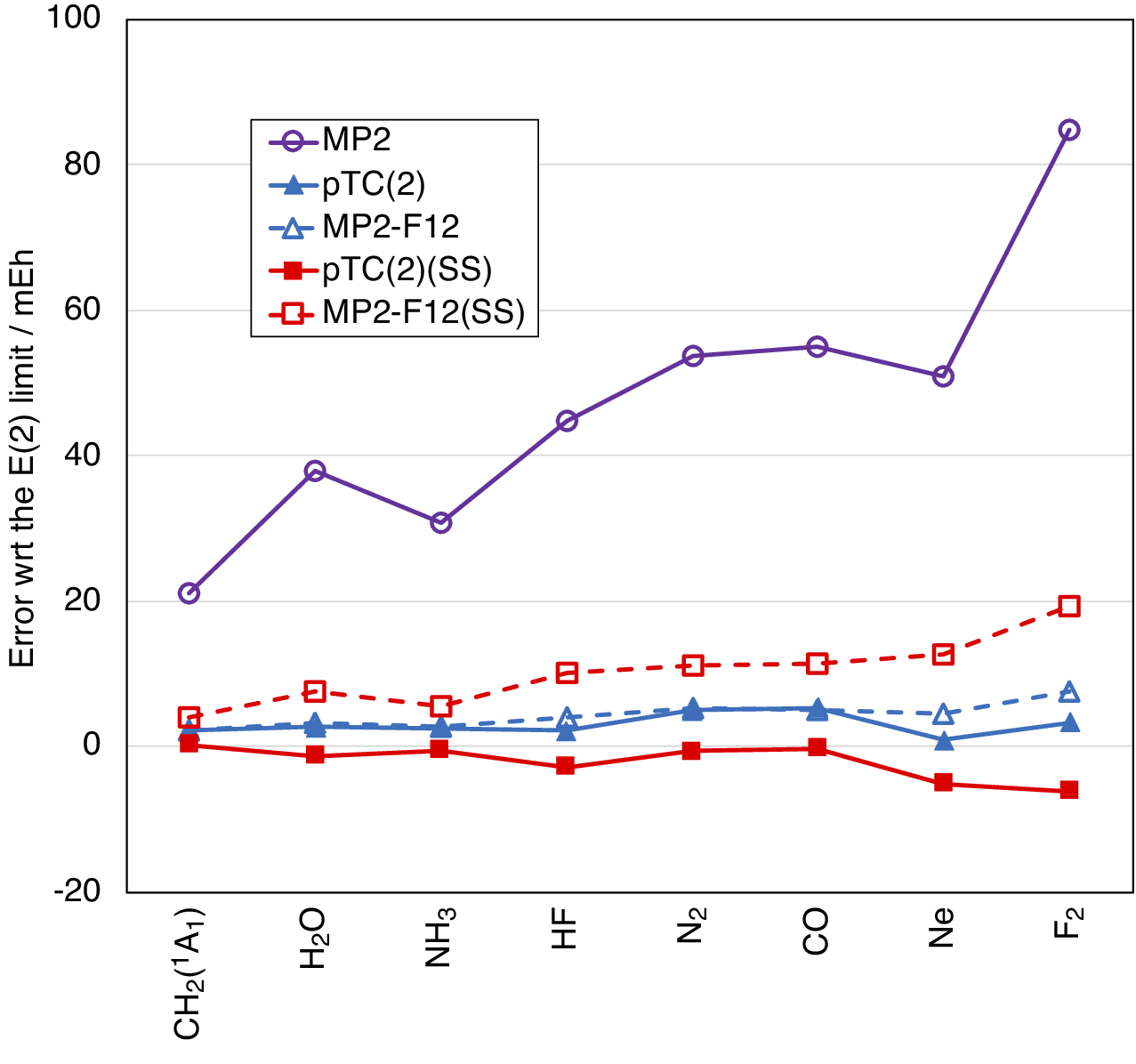}
	\caption{Errors of all-electron correlation energies in the aug-cc-pCVTZ basis set. The pTC(2) and MP2-F12 results are indicated by solid and dashed lines, respectively, and those with the SP and SS ans\"atze are displayed in blue and red.} \label{fig:acvtz}
\end{figure}
\begin{table}
\begin{center}
\caption
{\label{tab:val_avxz}
Statistical measures of valence-electron correlation energies (mE$_{\rm h}$) in aug-cc-pV{\it X}Z basis sets.}
\scalebox{1.00}{
\begin{tabular}{llrrrrr}
\hline\hline
&& \multicolumn{5}{c}{{\it X}} \\
\cline{3-7}
&& \multicolumn{1}{c}{D} & \multicolumn{1}{c}{T} & \multicolumn{1}{c}{Q} & \multicolumn{1}{c}{5} & \multicolumn{1}{c}{6} \\ 
\hline
$\bar\Delta$ & MP2 & 98.70 & 39.88 & 18.56 & 9.79 & 5.86 \\
& pTC(2) & 21.77 & 3.48 & 0.78 & 0.23 & 0.09 \\
& pTC(2)(SS) & 8.30 & -1.13 & -0.91 & -0.44 & -0.21 \\
& MP2-F12 & 12.71 & 3.89 & 1.46 & 0.58 & 0.23 \\
& MP2-F12(SS) & 18.61 & 9.06 & 3.80 & 1.62 & 0.71 \\
\hline
$\bar\Delta_{\rm abs}$ & pTC(2) & 21.77 & 3.48 & 0.80 & 0.26 & 0.12 \\
& pTC(2)(SS) & 11.00 & 2.96 & 1.03 & 0.47 & 0.22 \\
\hline
$\bar\Delta_{\rm cusp}$ & pTC(2) & -13.47 & -4.60 & -1.69 & -0.66 & -0.30 \\
& MP2-F12 & 5.90 & 5.17 & 2.34 & 1.04 & 0.47 \\
\hline
$\bar\Delta_{\rm opt}$ & SP ansatz & 9.78 & 2.49 & 0.84 & 0.39 & 0.16 \\
& SS ansatz & 14.96 & 10.20 & 4.71 & 2.05 & 0.92 \\
\hline\hline
\end{tabular}}
\end{center}
\end{table}
We compare pTC(2) with MP2-F12 for the test set molecules.
To assess the role of cusp conditions, we use the SS ansatz imposing the s-wave cusp condition both for singlet and triplet pairs,
\begin{align}
\hat {\mathcal R}^{\rm (SS)}_{12} = \frac{1}{2}f(r_{12}).
\end{align}
As an example, the errors of all-electron energies in aug-cc-pCVTZ are shown in FIG \ref{fig:acvtz}.
The pTC(2) energies resemble those of MP2-F12 quite well, while the differences become significantly large with the SS ansatz detariorating the triplet cusp condition.
The pTC(2)(SS) energies are always lower than pTC(2) with the SP ansatz in contradiction to the variational behavior of MP2-F12 $E_{\rm MP2\mathchar`-F12(SS)}>E_{\rm MP2\mathchar`-F12}$.

We detail the discussion employing the statistical measures, the mean error $\bar\Delta$, the mean absolute error $\bar\Delta_{\rm abs}$ for the non-variational pTC, the mean deviation of the SS energy from SP to indicate the error due to the p-wave cusp condition,
\begin{align}
\bar\Delta_{\rm cusp} = \overline {E_{\rm SS}-E_{\rm SP}},
\end{align}
and the mean absolute deviation of pTC(2) from MP2-F12 for the optimality of the geminal ansatz,
\begin{align}
\bar\Delta_{\rm opt} = \overline {|E_{\rm pTC}(2)-E_{\rm MP2\mathchar`-F12}|}.
\end{align}
The measures for the valence- and all-electron correlation energies are given in Tables \ref{tab:val_avxz} and \ref{tab:all_acvxz}.
In contrast to the others, $\bar\Delta$ of pTC(2)(SS) converges from below except for the valence correlation in aug-cc-pVDZ, where the first-order $\langle \hat {\mathcal H}_{\rm 2h} + {\mathcal H}_{\rm 2v}\rangle$ makes large positive contributions for several systems.
$\bar\Delta_{\rm cusp}$ is always negative for pTC(2) contrary to MP2-F12.
Accordingly, pTC(2)(SS) contains error cancelation between the deficiency of the triplet cusp condition and basis set incompleteness with negative and positive contributions, respectively.
The optimality index $\bar\Delta_{\rm opt}$ of the SP ansatz is several times as small as the SS one, indicating the importance of the cusp conditions for proper transcorrelation.

\begin{figure}[b]
	\includegraphics[width = 26em]{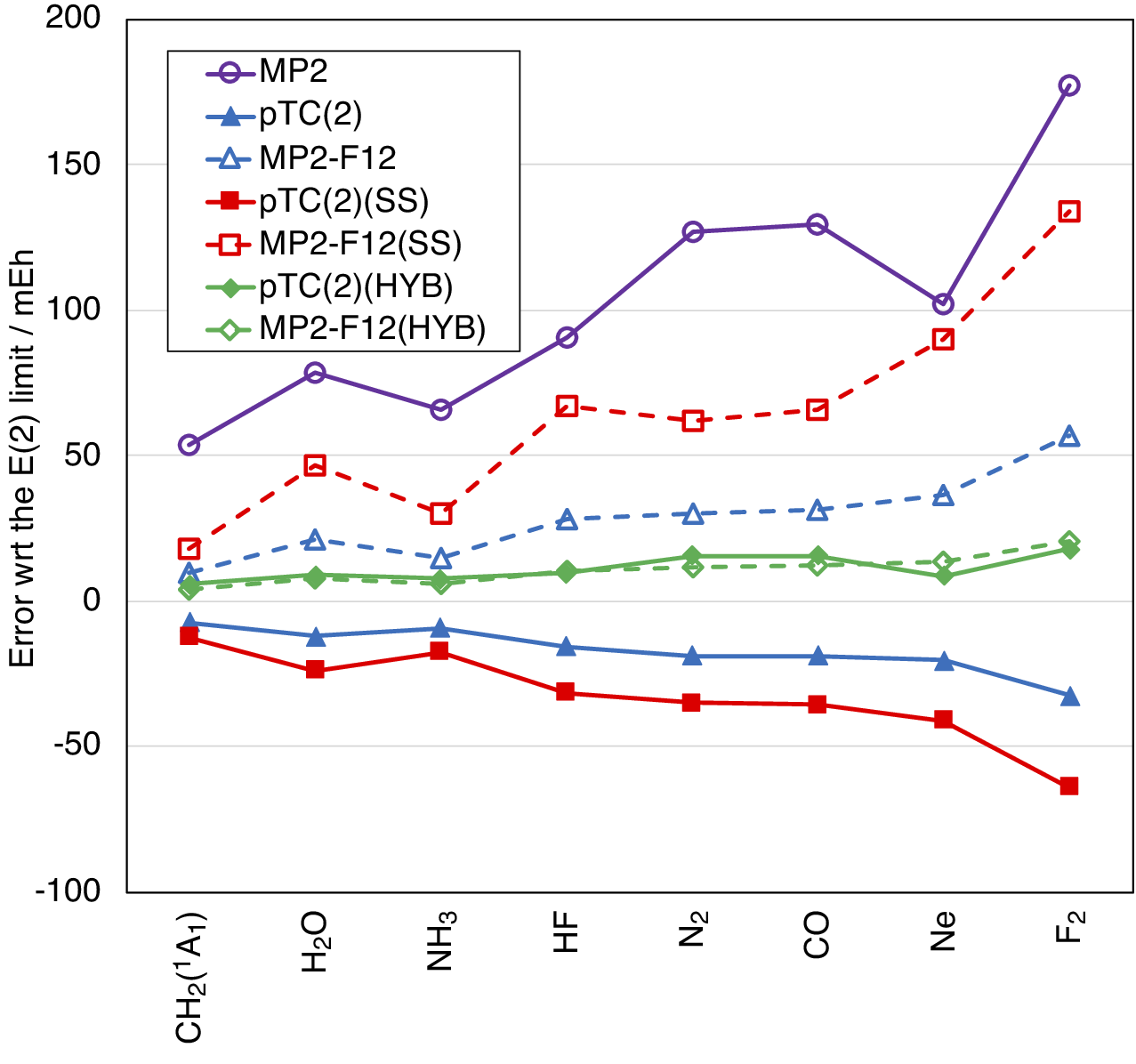}
	\caption{Errors of all-electron correlation energies in the aug-cc-pVTZ basis set. The expressions are same as those in FIG. \ref{fig:acvtz} except for the HYB result indicated in green.} \label{fig:avtz}
\end{figure}

We finally discuss the treatment of core-valence correlation energies.
FIG. \ref{fig:avtz} exhibits the errors of all-electron correlation energies of H$_2$O in aug-cc-pVTZ.
Comparred to the result with aug-cc-pCVTZ in FIG. \ref{fig:acvtz}, the errors increase significantly in the absence of the core functions both for pTC and MP2-F12.
This deterioration gives negative and positive impacts for pTC(2) and MP2-F12, respectively.
MP2-F12(HYB) markedly improves the error in the CC and CV correlation energies.
 pTC(2)(HYB) employs the CC and CV geminal amplitudes of MP2-F12 to show a similar performance.
\begin{table}
\begin{center}
\caption
{\label{tab:all_acvxz}
Statistical measures of all-electron correlation energies (mE$_{\rm h}$) in aug-cc-pCV{\it X}Z basis sets.}
\scalebox{1.00}{
\begin{tabular}{llrrrrr}
\hline\hline
&& \multicolumn{5}{c}{{\it X}} \\
\cline{3-7}
&& \multicolumn{1}{c}{D} & \multicolumn{1}{c}{T} & \multicolumn{1}{c}{Q} & \multicolumn{1}{c}{5} & \multicolumn{1}{c}{6} \\ 
\hline
$\bar\Delta$ & MP2 & 127.64 & 47.37 & 20.79 & 10.85 & 6.43 \\
& pTC(2) & 15.25 & 3.00 & 0.88 & 0.24 & 0.11 \\
& pTC(2)(SS) & -6.41 & -2.05 & -0.55 & -0.32 & -0.14 \\
& MP2-F12 & 22.19 & 4.31 & 0.96 & 0.25 & 0.08 \\
& MP2-F12(SS) & 47.74 & 10.22 & 2.66 & 0.85 & 0.33 \\
\hline
$\bar\Delta_{\rm abs}$ & pTC(2) & 15.40 & 3.00 & 0.88 & 0.24 & 0.11 \\
& pTC(2)(SS) & 14.35 & 2.10 & 0.58 & 0.34 & 0.15 \\
\hline
$\bar\Delta_{\rm cusp}$ & pTC(2) & -21.66 & -5.06 & -1.43 & -0.56 & -0.25 \\
& MP2-F12 & 25.56 & 5.91 & 1.70 & 0.60 & 0.25 \\
\hline
$\bar\Delta_{\rm opt}$ & SP ansatz & 10.01 & 1.35 & 0.16 & 0.09 & 0.04 \\
& SS ansatz & 54.16 & 9.90 & 3.21 & 1.18 & 0.47 \\
\hline\hline
\end{tabular}}
\end{center}
\end{table}
\begin{table}
\begin{center}
\caption
{\label{tab:all_avxz}
Statistical measures of all-electron correlation energies (mE$_{\rm h}$) in aug-cc-pV{\it X}Z basis sets.}
\scalebox{1.00}{
\begin{tabular}{llrrrrr}
\hline\hline
&& \multicolumn{5}{c}{{\it X}} \\
\cline{3-7}
&& \multicolumn{1}{c}{D} & \multicolumn{1}{c}{T} & \multicolumn{1}{c}{Q} & \multicolumn{1}{c}{5} & \multicolumn{1}{c}{6} \\ 
\hline
$\bar\Delta$ & MP2 & 178.88 & 103.12 & 60.76 & 43.23 & 31.56 \\
& pTC(2) & 2.05 & -16.82 & -14.92 & -11.59 & -8.12 \\
& pTC(2)(SS) & -26.89 & -32.67 & -22.33 & -15.34 & -9.97 \\
& pTC(2)(HYB) & 34.74 & 11.20 & 5.24 & 3.74 & 2.87 \\
& MP2-F12 & 44.54 & 28.66 & 17.85 & 12.50 & 8.28 \\
& MP2-F12(SS) & 91.74 & 64.30 & 36.86 & 23.65 & 14.36 \\
& MP2-F12(HYB) & 23.45 & 10.81 & 5.81 & 4.07 & 3.02 \\
\hline
$\bar\Delta_{\rm abs}$ & pTC(2) & 9.24 & 16.82 & 14.92 & 11.59 & 8.12 \\
& pTC(2)(SS) & 27.01 & 32.67 & 22.33 & 15.34 & 9.97 \\
& pTC(2)(HYB) & 34.74 & 11.20 & 5.24 & 3.74 & 2.87 \\
\hline
$\bar\Delta_{\rm opt}$ & SP ansatz & 42.47 & 45.48 & 32.77 & 24.09 & 16.40 \\
& SS ansatz & 118.63 & 96.98 & 59.19 & 38.99 & 24.33 \\
& HYB ansatz & 11.97 & 2.63 & 0.80 & 0.37 & 0.17 \\
\hline\hline
\end{tabular}}
\end{center}
\end{table}
The statistical measures of the corresponding calculations for the test set molecules are listed in Table \ref{tab:all_avxz}.
The mean errors of MP2-F12(HYB) and pTC(2)(HYB) across 10 mE$_{\rm H}$ from {\it X=}T to Q.
The optimality index $\bar\Delta_{\rm opt}$ of the HYB ansatz is from 20 to 100 times as small as the SP one for $X \ge$T, suggesting the possibility of diverting geminal amplitudes from low-order F12 theory to higher-order pTC calculations.
Nevertheless, the very sharp convergence as in Table \ref{tab:all_acvxz} is not attained solely by the optimization of amplitudes, and augmentations with core functions appear to be the best method for highly accurate core-valence correlation energies.

\section{CONCLUSIONS}\label{sec:conclusions}
We have introduced a projective transcorrelation inspired by the F12 ansatz.
The effective Hamiltonian $\hat {\mathcal H}_{\rm pTC}$ is universal, spin-free, satisfying the s- and p-wave cusp conditions simultaneously, and terminates at the four-body interaction.
In the SR case, the second order contribution of $\hat {\mathcal H}_{\rm pTC}$ is equivalent to the V-term of MP2-F12 plus additional terms not present in MP2-F12 with GBC and $\hat{\mathcal Q}^{\rm (SO)}_{12}$.
Higher-order contribution can be automatically incorporated by involving $\hat {\mathcal H}_{\rm pTC}$ in correlated calculations.
Numerical results of pTC(2) indicate that single contractions through the complementary projector $\hat Q_n=1-\hat P_n$ decay very rapidly with the basis set size.
This suggests that $\hat {\mathcal H}_{\rm pTC}$ can be practically approximated by omitting the three- and four-body operators, $\hat {\mathcal H}_{\rm 3q}$ and $\hat {\mathcal H}_{4}$ quadratic to $\hat Q_n$ as
\begin{align}
\hat {\mathcal H}_{\rm pTC}\approx \hat {\mathcal H} + \hat {\mathcal H}_{\rm 2h}+\hat {\mathcal H}_{\rm 2w}+\hat {\mathcal H}_{\rm 3l},
\end{align}
in accord with common approximations of four-electron integrals in F12 theory, which makes the implementation extremely easy. 
These features are significant advantages over the previous TC of Boys and Handy.

Due to the non-hermitian nature of the effective Hamiltonian, we need to pay special attention to the accuracy of transcorrelation.
The deficiency of cusp conditions and basis set incompleteness can contribute with opposite signs, leading to error cancelation.
The cusp conditions play an important role in transcorrelation, and the validity of the ansatz must be confirmed through comparisons with variational calculations.
The pTC is universal and will provide useful applications when combined with FCI solvers.
We plan to report on the integration of pTC with promising solvers for strongly correlated electrons, including selected coupled-cluster\cite{xu2018full,xu2020towards} and model space QMC.\cite{ten2013stochastic,ten2017multi}

\section*{ACKNOWLEDGEMENT}
This work is partially supported by the Grant-in-Aids for Scientific Research (A) (Grant No. 22H00316) from the Japan Society for the Promotion of Science (JSPS) and MEXT as ``Program for Promoting Researches on the Supercomputer Fugaku'' (Realization of innovative light energy conversion materials, Grant Number JPMXP1020210317).

\section*{AUTHOR DECLARATIONS}
\subsection*{Conflict of Interest}
The authors have no conflicts to disclose.

\subsection*{Author Contributions}
Seiichiro Ten-no: Conceptualization (equal); Data curation (equal); Formal analysis (equal); Funding acquisition (equal); Investigation (equal); Methodology (equal); Project administration (equal); Resources (equal); Software (equal); Supervision (equal); Validation (equal); Visualization (equal); Writing -- original draft (equal); Writing -- review \& editing (equal).

\section*{DATA AVAILABILITY}
The data that support the findings of this study are available from the corresponding author upon reasonable request.

\section*{REFERENCES}
\bibliography{refs}
\end{document}